\documentclass{vgtc}                          %
\usepackage{graphicx}
\graphicspath{{figures/}{pictures/}{images/}{./}} %

\usepackage{times}                     %
\usepackage{tabu}                      %
\usepackage{booktabs}                  %
\usepackage{lipsum}                    %
\usepackage{mwe}                       %
\usepackage{titlesec}
\usepackage{enumitem}
\titlespacing*{\section}{0pt}{6pt}{4pt}
\titlespacing*{\subsection}{0pt}{4pt}{2pt}

\usepackage{mathptmx}                  %

\usepackage{xcolor}
\usepackage{xspace}
\usepackage{enumitem}

\usepackage{tabularx} 
\usepackage{wrapfig}  
\usepackage{balance}

\newcommand{\tool}[0]{UNIPO\xspace}

\newcommand{\paraheader}[1]{\par\vspace{-\parskip}\vspace{4pt}\noindent\textbf{#1}}

\definecolor{green}{HTML}{57B069}
\definecolor{pink}{HTML}{ED8FA6}
\definecolor{softBlue}{HTML}{407CBF}

\onlineid{0}

\vgtccategory{Systems and Applications}

\vgtcinsertpkg

\title{\tool{}: Unified Interactive Visual Explanation for\\ RL Fine-Tuning Policy Optimization}

\author{%
  Aeree Cho$^1$ \and
  Alexander D. Greenhalgh$^1$ \and
  Jonathan Bodea$^1$ \and
  Anthony Peng$^1$ \and
  Duen Horng (Polo) Chau\thanks{$^{1}$Georgia Tech.\\
  \{%
  \href{mailto:aeree@gatech.edu}{aeree} $|$
  \href{mailto:agreenhalgh3@gatech.edu}{agreenhalgh3} $|$
  \href{mailto:jbodea6@gatech.edu}{jbodea6} $|$
  \href{mailto:speng65@gatech.edu}{speng65} $|$
  \href{mailto:polo@gatech.edu}{polo}%
  \}@gatech.edu}%
}

\teaser{
  \centering
  \includegraphics[width=\linewidth]{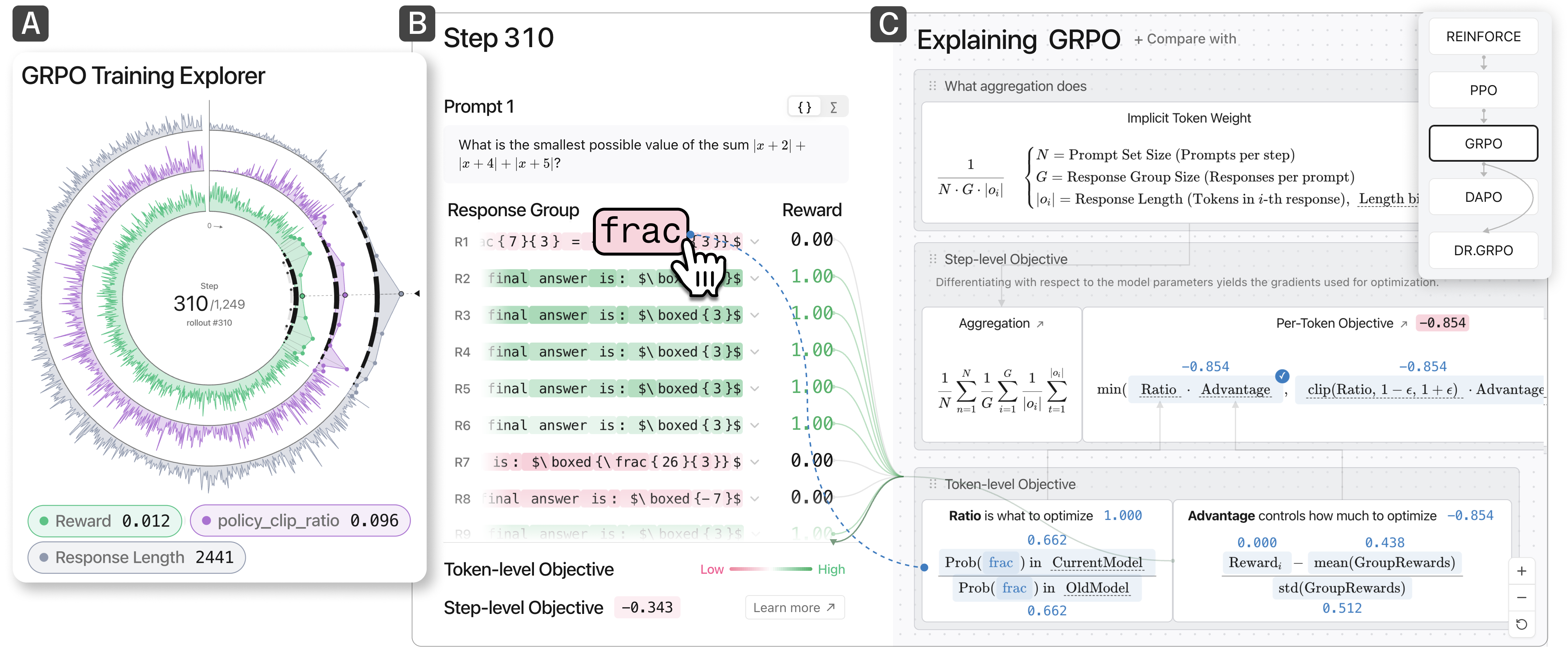}
  \caption{\tool{} unifies the visual explanation of policy optimization algorithms for RL fine-tuning through three coordinated views that progressively reveal how algorithmic design shapes training.
  \textbf{(A) Training Explorer} summarizes the full training run with a multi-metric radial plot.
  \textbf{(B) Step Inspector} visualizes token-level objectives for a selected step's  prompt-responses set, revealing which tokens are \textcolor{green}{reinforced} or \textcolor{pink}{suppressed} in the next gradient update. 
  \textbf{(C) Algorithm Explainer} decomposes the active objective, mapping values computed for the selected token to bridge abstract notation with real training behavior.}
  \label{fig:teaser}
}

\abstract{
    Reinforcement learning has emerged as a dominant technique for fine-tuning the behavior of large language models, with policy optimization (PO) algorithms such as GRPO, DAPO, and Dr. GRPO emerging in rapid succession to advance state-of-the-art reasoning and alignment performance. However, the modular differences between these algorithms, including targeted improvements to clipping, advantage estimation, and reward aggregation, are introduced across separate papers with inconsistent notation, making them difficult to compare and intimidating to the non-expert community. We present \tool{}, to our knowledge the first interactive visualization tool that exposes the token-level training dynamics of RL fine-tuning algorithms through a unified design. \tool{} connects three complementary views, a high-level training overview, a step-level prompt and response inspector, and a side-by-side algorithm comparison, allowing learners to observe how individual design decisions propagate through training. Through two usage scenarios, we demonstrate how \tool{} supports both classroom instruction for non-experts and algorithm selection for AI practitioners. Our tool is open-source and publicly available at {\url{https://poloclub.github.io/unipo/}}.
} %


\begin{document}

\firstsection{Introduction}

\maketitle

Reinforcement Learning fine-tuning has emerged as a dominant technique for aligning pre-trained large language models with task-specific goals, including chain-of-thought reasoning, math solving, and persona matching \cite{RLHF_christiano2023_deepreinforcementlearninghuman, bai2022traininghelpfulharmlessassistant,RLHF_ouyang2022traininglanguagemodelsfollow}. Building on policy optimization (PO) algorithms like REINFORCE and PPO, derivatives such as GRPO, DAPO, and Dr. GRPO have emerged in rapid succession, each making targeted improvements to the shared PO framework. Despite their widespread adoption in production-grade LLMs \cite{Guo_2025,DAPO_yu2025}, the internal mechanisms driving these algorithms' performance remain difficult to build intuition for, particularly for learners and AI practitioners new to RL fine-tuning, requiring deep mathematical expertise and careful comparison across a growing body of work.

Enabling better understanding of these algorithms is critical to expanding RL fine-tuning's accessibility to expert and non-expert audiences. Existing resources, including books \cite{sutton_2018,lambert2026reinforcementlearninghumanfeedback}, blog posts \cite{lambert2022illustrating}, and video tutorials \cite{kilcher2025grpo} are non-interactive, obscuring how each algorithm's modifications shape training behavior.
Monitoring tools used by practitioners, like Weights \& Biases \cite{wandb}, present a large amount of 
training metrics, making it difficult to isolate how individual algorithmic mechanisms contribute to training behavior.
Comparative blog posts \cite{zhang2025grpo} and benchmarking studies \cite{lian2025comparativeanalysisparametrictuning} provide direct algorithmic comparisons but cannot communicate how modular design decisions propagate through training at the token level.
Interactive visualization is necessary to support hands-on exploration of training dynamics and side-by-side algorithmic comparison, particularly for non-experts who lack the background to reconstruct algorithm behavior from mathematical notation. To address this gap, we contribute:
\begin{enumerate}
[nosep,leftmargin=*]
  \item \textbf{\tool{}, a web-based interactive tool for unified visual explanation of five RL fine-tuning algorithms}~(\autoref{fig:teaser}). \tool{} visualizes foundational REINFORCE and PPO, and state-of-the-art GRPO, DAPO, and Dr. GRPO. Three coordinated views, \textit{Training Explorer}, \textit{Step Inspector}, \textit{Algorithm Explainer}, reveal training dynamics from high-level metrics down to token-level objective computations. This layered design allows non-experts to build algorithmic intuition before engaging with the underlying mathematics, while also supporting AI practitioners who seek to compare the objective functions of these algorithms. 
  \item \textbf{A novel visualization that modularizes mathematical notation across RL fine-tuning algorithms into shared structural components}. The \textit{Algorithm Explainer} decomposes each algorithm's objective into a standardized set of components: aggregation, per-token objective, and constraints. This modular decomposition enables a consistent visual representation that precisely highlights cross-algorithm differences, allowing users to compare any two models within the RL fine-tuning family.
  \item \textbf{An extensible, open-source implementation supporting user-provided training logs and algorithm definitions.} The RL fine-tuning landscape continues to expand, and \tool{} is designed to seamlessly accommodate new training data and algorithms through a unified interface.
  Users can visualize their own training runs and register new algorithms beyond the five presets using our canonical JSON schema. The schema and open-source implementation are available at {\url{https://github.com/poloclub/unipo}}, enabling \tool{} to grow alongside the evolving RL fine-tuning ecosystem.

\end{enumerate} 

\section{Related Work}

\paraheader{Policy Optimization for LLM Fine-Tuning.} Reinforcement Learning
is now a primary 
technique in LLM fine-tuning, both for aligning models with human preferences (RLHF \cite{RLHF_ouyang2022traininglanguagemodelsfollow}) and for improving reasoning through verifiable rewards \cite{lambert2025tulu3pushingfrontiers}.
Within RL, policy optimization (PO), which directly optimizes the model from reward signals, has become the dominant family of algorithms \cite{wang2026reinforcementlearningllmposttraining}. 
Inspired by the REINFORCE algorithm \cite{REINFORCE_1992}, Proximal Policy Optimization (PPO) \cite{PPO_schulman2017} became the dominant approach for LLM alignment by introducing a learned critic to evaluate the policy at each step, at the cost of training a second model. GRPO \cite{GRPO_shao2024deepseekmathpushinglimitsmathematical} eliminates this critic by
computing a group-relative advantage from $G$ parallel responses. 
This enabled DeepSeek-R1 \cite{Guo_2025} to match larger models, establishing GRPO as a dominant RL fine-tuning algorithm. 
DAPO \cite{DAPO_yu2025} refines GRPO with per-token importance ratios, asymmetric clipping, and the removal of the KL penalty, reaching state-of-the-art performance on their benchmark in half of GRPO's training steps. 
The Dr. GRPO \cite{drGRPO_liu2025} preserves GRPO's core structure while normalizing response length to stabilize performance on longer responses. 
Each of these algorithms refines its predecessor but is introduced in its own paper with distinct notation, making direct comparison difficult \cite{zhang2025grpo,wang2025lambdagrpounifyinggrpoframeworks}.

\paraheader{Visualization Tools for ML Explainability.} A growing body of research has created interactive visualization tools that ensure foundational machine learning models remain approachable to non-experts \cite{smilkov2017directmanipulationvisualizationdeepnetworks,hohman2018visualanalyticsdeeplearning,liu2024techniques}.  The CNN Explainer \cite{Wang_2021} establishes a progressive design pattern, allowing users to explore model mechanisms at varying depths through explicit visualization of all model inputs, intermediate steps, and outputs. The Transformer Explainer \cite{transformerExplainer_CHI} and Transforlearn \cite{gao2023transforlearn} 
extends this approach with in-browser inference, exposing attention and token probabilities. Other tools focus on inspecting specific components of trained models, such as attention patterns~\cite{vig-2019-multiscale,attentionViz_2024} and multimodal representations~\cite{VLInterpret_2022}, and reasoning in question answering~\cite{shao2023visual}. These tools focus on inference-time behavior, not the training dynamics. 
A separate line of work visualizes RL agent behavior in interactive environments \cite{hu2024interactive, agarwal2020bombalytics, agarwal2020visualizing}, but focuses on an agent's actions rather than how the underlying policy is updated.
Practitioner tools such as Weights \& Biases or MLflow~\cite{Zaharia2018AcceleratingTM} visualize these training dynamics, but abstract away algorithmic mechanisms in favor of convergence metrics, requiring model implementation and GPU access -- both out of reach for non-experts. 
% To our knowledge, no existing tool exposes the hierarchical breakdown of how these RL fine-tuning algorithms derive their model updates through a unified token-level visualization, nor contextualizes each algorithm's contribution through the evolution of the RL fine-tuning algorithm family.
To our knowledge, no existing tool exposes how these algorithms derive model updates through a unified token-level visualization, nor contextualizes each within the evolution of the RL fine-tuning family.

\section{Design Goals}\label{sec:design-goal}
% Through reviewing related works ~\cite{hohman2018visualanalyticsdeeplearning, Wang_2021,transformerExplainer_CHI} and challenges of communicating RL fine-tuning algorithms to both non-expert and expert audiences, we establish three primary design goals:
\tool{} primarily targets \textit{RL learners}, who build intuition for how design shapes training,
especially students in a classroom, since \tool{} is a web app running on each learner's device without specialized hardware (\autoref{sec:learning_grpo}), and \textit{practitioners}, who compare algorithms before a costly run (\autoref{sec:algorithm_comparison}). 
We synthesize their needs from prior ML-explainability tools~\cite{hohman2018visualanalyticsdeeplearning, Wang_2021,transformerExplainer_CHI} and the difficulty of reconciling RL algorithms across inconsistent notation~\cite{zhang2025grpo}, mapping them to three design goals:

\begin{enumerate}[topsep=1pt, itemsep=0mm, parsep=1pt, leftmargin=*, label=\textbf{G\arabic*.}, ref=\textbf{G\arabic*}]
\item  \textbf{High-level overview of RL fine-tuning training dynamics.}\label{goal:overview} A single RL fine-tuning run produces thousands of training steps across multiple metrics, including reward, loss, and KL divergence. Non-expert users need an approachable entry point to orient themselves to these high-level training dynamics before engaging with token-level mechanics \cite{Schneiderman_TheEyesHaveIt,Wang_2021}. Monitoring tools like Weights \& Biases \cite{wandb} are intimidating to a non-experts and abstract away the underlying mathematics from AI practitioners \cite{hohman2018visualanalyticsdeeplearning}. \tool{} addresses both audiences through a radial training overview that orients non-experts to the training run while offering practitioners a navigable entry point into specific training steps. 

\item  \textbf{Expose the token-level computations that connect response groups to gradient updates.}\label{goal:token-level} The path from a sampled response group to a gradient update passes through token-level importance ratios and advantages that are obscured in static-text resources. Without observing how these values arise from real training runs, users cannot build intuition for why specific ``correct'' responses are reinforced and others are suppressed \cite{smilkov2017directmanipulationvisualizationdeepnetworks}. Non-experts need straightforward explanations of these computations, while AI practitioners need to connect token-level behavior back to the mathematical notation of each algorithm's objective function \cite{Endert_2017}. \tool{} bridges this gap by surfacing the token-level importance ratios and advantages that drive each gradient update, anchoring the mathematical notation to the concrete values produced during training.

\item  \textbf{Contextualize each algorithm within the evolution of the RL fine-tuning family.}\label{goal:algorithm} State-of-the-art RL fine-tuning algorithms have evolved from targeted improvements to their predecessor's weaknesses: GRPO removes PPO's critic model, DAPO introduces dynamic sampling and asymmetric clipping, and Dr. GRPO refines normalization for long responses. Without seeing how each algorithm builds on its predecessors, users cannot contextualize the family's development \cite{Lakatos1976}. Existing static-text resources require users to manually reconcile notation across literature, time-consuming for experts~\cite{shengyi2022the37implementation} and intimidating for newcomers. \tool{} positions each algorithm within the family's progression, supporting cross-algorithm comparison that renders any two algorithms' objective functions side-by-side.

\end{enumerate}

\section{System Design and Implementation}
To operationalize the three design goals from \autoref{sec:design-goal},
\tool{} utilizes three distinct connected views: \textit{Training Explorer} (\autoref{sec:training-explorer}), \textit{Step Inspector} (\autoref{sec:step-inspector}), and \textit{Algorithm Explainer} (\autoref{sec:algorithm-explainer}). These views progressively increase in complexity, from a high-level training overview, to step-level response groups, to a mathematical decomposition of the active algorithm's objective function. This layered design lets non-experts build intuition with the training dynamics before approaching the underlying mathematics, while practitioners move directly to comparing objective functions at the appropriate mathematical depth.

% This initial release focuses on five representative PO methods: REINFORCE, PPO, GRPO, DAPO, and Dr. GRPO. Using OpenRLHF, we fine-tune Llama-3.2-1B-Instruct with each of these five algorithms on a math reasoning task from the MATH dataset ($\sim$1$,$000 steps per run). The frontend is built with Svelte and D3, loading a unified JSON training schema to visualize training dynamics directly in the browser.
Built with Svelte and D3, \tool{} is a browser-based frontend that reads a run from a single canonical JSON schema carrying per-step, token-level logs; any framework that emits this schema can drive \tool{}. 
% Logging granularity is thus left to the user, and large runs can be reduced to representative steps or samples. 
Since \tool{} only reads these logs, granularity is user-defined: large runs can log only representative steps or samples (in our OpenRLHF runs, token-level logging added negligible, $<$1\%, overhead).
For responsiveness, we recommend chunking logs per step on a content delivery network (CDN) and registering their base URL with \tool{}, which then fetches only the inspected step. 
We provide five presets---REINFORCE, PPO, GRPO, DAPO, and Dr.\ GRPO---built from Llama-3.2-1B-Instruct runs on MATH.
Since \tool{} consumes only the exported schema, the same views extend in principle to other models, tasks, and reward types; validating this transfer is a clear direction for future work.

\subsection{Training Explorer}
\label{sec:training-explorer}

The \textit{Training Explorer} provides a high-level overview of the training dynamics across all steps (each corresponds to one gradient update) of the training run (\ref{goal:overview}, \autoref{fig:teaser}A). Training metrics are visualized through a radial plot, where each concentric ring represents a metric (e.g., reward, policy loss, KL divergence, clipping behavior) corresponding to the training step's angular position.
We use a radial layout to present the entire training run as a single compact glyph, aligning multiple metrics on a shared angular axis so they can be compared at a given step along one spoke \cite{draper2009survey}. It also maps a run of any length into one fixed circular footprint, making it easier to compare the trajectories of different runs as whole-shape glyphs at a constant size \cite{borgo2013glyph}.
To maintain readability at scale, we apply LTTB (Largest-Triangle-Three-Buckets) downsampling \cite{steinarsson2013downsampling} to reduce clutter while preserving the overall shape and key features of the data.
A fisheye interaction locally expands dense regions of the radial plot, enabling users to inspect and select individual steps while preserving the global trajectory.
Users can overlay up to four metrics to view the full training dynamics of each algorithms. 

\subsection{Step Inspector}
\label{sec:step-inspector}

\begin{figure}[t]
  \centering
  \includegraphics[width=\linewidth]{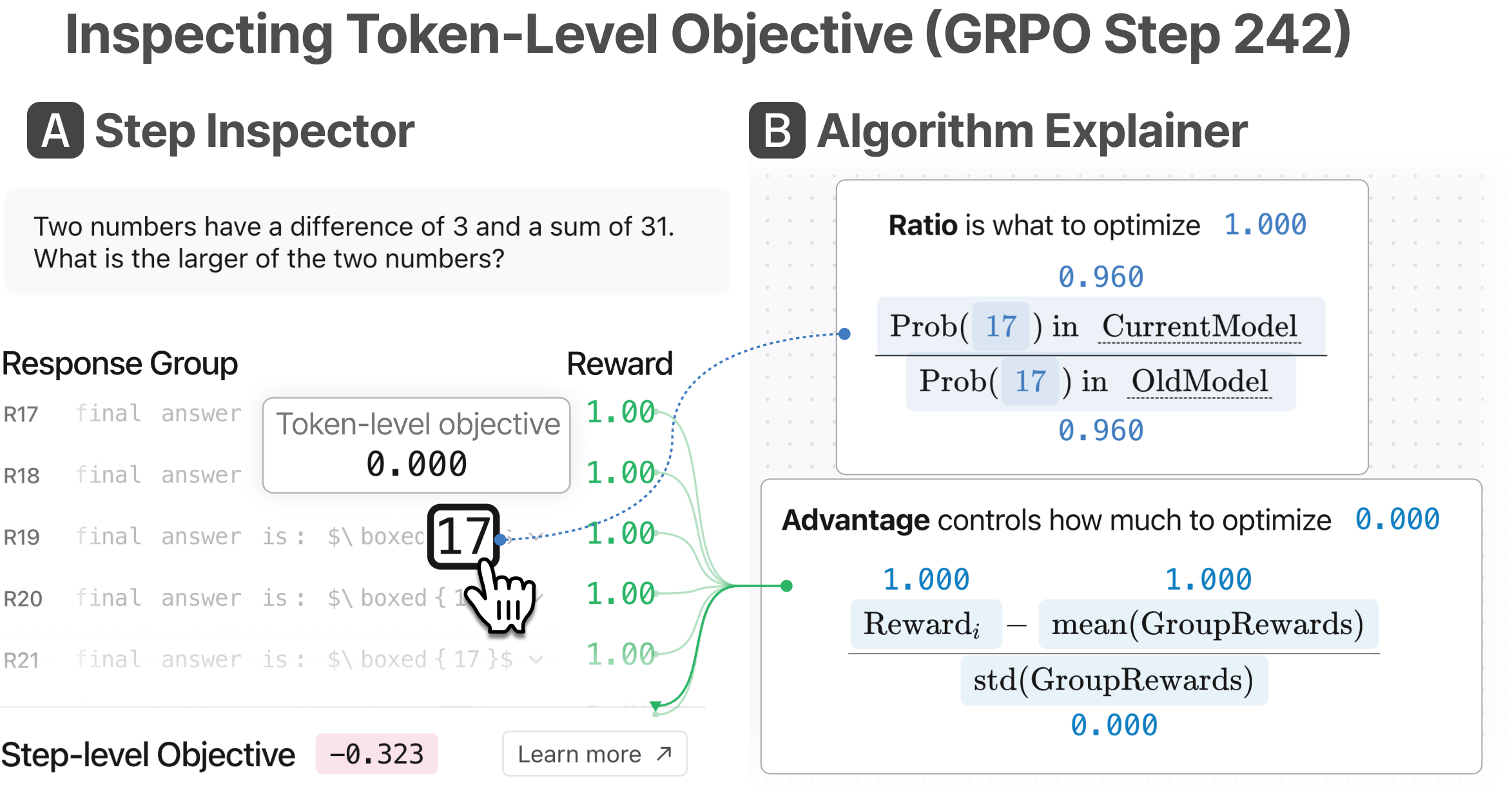}
  \caption{Selecting a token in the \textbf{(A) Step Inspector} opens its computation in the \textbf{(B) Algorithm Explainer}. 
  Here, ``17'' receives a token-level objective of $0.000$ even though the reward is \textcolor{green}{$1.00$}. Every response in the group is correct, so the \textit{Advantage} collapses to \textcolor{softBlue}{$0.000$}. This reveals that GRPO reinforces responses relative to the group, not by correctness alone.}
  \label{fig:step_inspector}
\end{figure}

The \textit{Step Inspector} presents the prompt-response group at a selected training step, 
along with the step-level aggregated objective and its underlying token-level contributions, including rewards and gradients. 
Responses are visualized as token sequences, with token-level objective values encoded as a color overlay on the text.
Lower values are shown in \textcolor{pink}{pink}, indicating the model will decrease the probability of those tokens in the next update, while higher values are shown in \textcolor{green}{green}, indicating the opposite (\autoref{fig:teaser}B).

While response-level rewards provide a high-level signal of preference, gradient updates in RL fine-tuning are computed at the token level through importance ratios and advantages. By exposing token-level objective values, the \textit{Step Inspector} makes explicit how each token contributes to the final gradient update (\autoref{fig:teaser}B and \autoref{fig:step_inspector}A), directly linking sampled responses to parameter updates (\ref{goal:token-level}). Users can select individual tokens to open the \textit{Algorithm Explainer} (\autoref{sec:algorithm-explainer}), which renders the objective function with the numerical values used to compute the selected token's gradient. This interaction bridges abstract formulations with the implementation-level quantities produced during training.

\subsection{Algorithm Explainer}
\label{sec:algorithm-explainer}

The \textit{Algorithm Explainer} uses a modular representation that enables cross-algorithm comparison and contextualizes the active algorithm within the evolution of the RL fine-tuning family (\ref{goal:algorithm}). 
We decompose each step-level objective into \textit{aggregation} terms, \textit{per-token objective} terms, and additional \textit{constraints}, and further break down each \textit{per-token objective} into the \textit{optimization target}, such as the importance ratio, and the \textit{optimization strength}, such as the advantage (\autoref{fig:teaser}C). Because these components recur across RL fine-tuning algorithms, this shared structure enables direct comparison while preserving their specific differences.

The view operationalizes this representation by presenting the active algorithm's objective function at multiple levels of complexity. Users can click each component to open an explanation card that connects abstract objective terms with the concrete computations performed at a training step.
The \textit{Token-level Objective Card} expands the per-token objective into computations such as the importance ratio and advantage. The \textit{Aggregation Card} traces how these token-level values are pooled across response groups and how they shape the resulting token-level gradients.
Hovering over individual terms reveals tooltip explanations that make the mathematical notation accessible to non-expert learners.

\begin{figure}[t]
  \centering
  \includegraphics[width=0.85\linewidth]{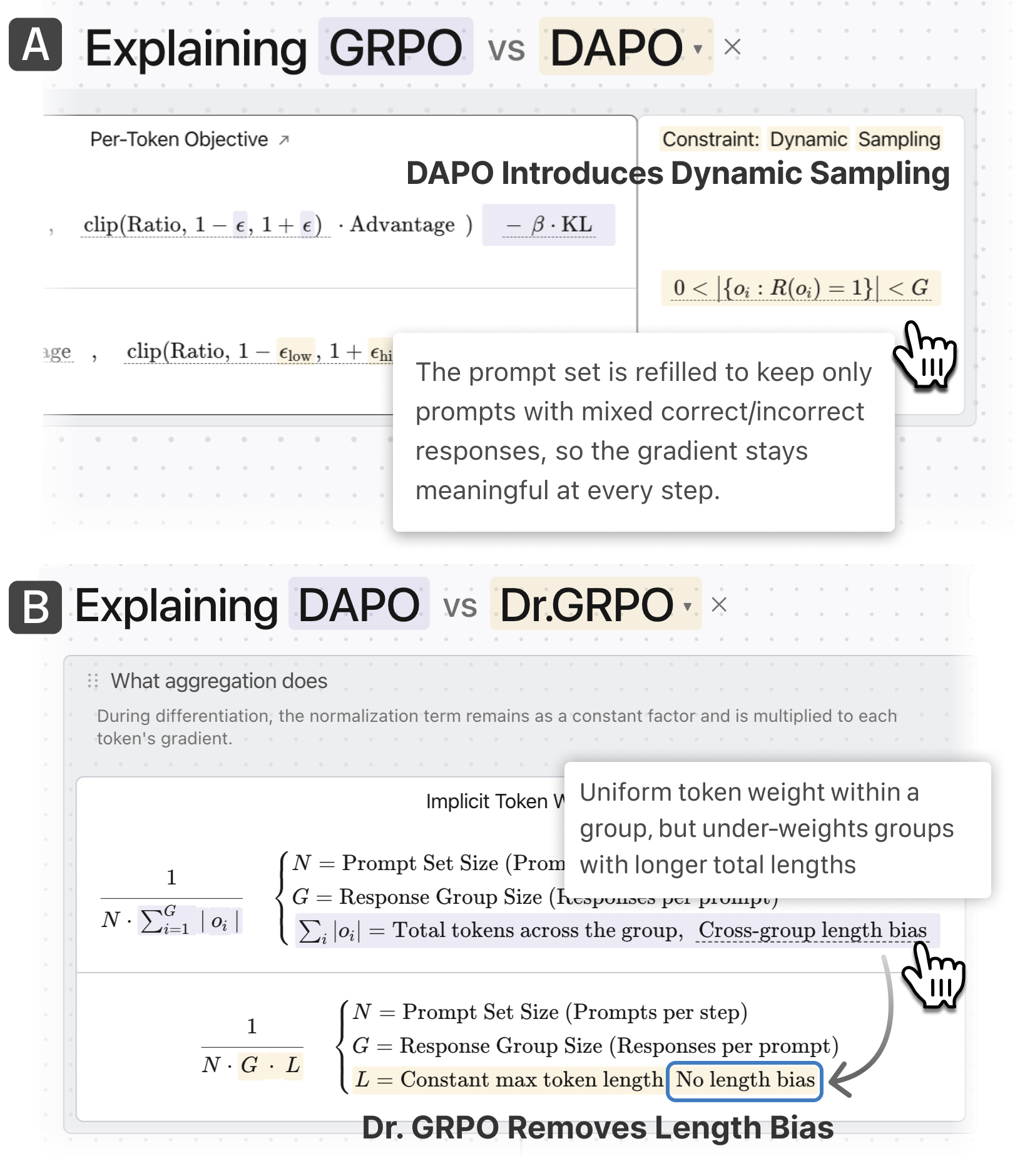}
\caption{\textit{Algorithm Explainer}'s \textbf{Comparison} mode renders two algorithms side-by-side with color-coded differences, revealing evolutionary relationships across policy optimization methods. 
\textbf{(A)} GRPO vs. DAPO surfaces DAPO's added \textit{Dynamic Sampling} constraint, with tooltip explaining it in plain language for non-experts. 
\textbf{(B)} DAPO vs. Dr.\ GRPO contrasts aggregation strategies, annotating DAPO's cross-group length bias and Dr.\ GRPO's bias-free normalization.}
  \label{fig:algorithm_explainer}
\end{figure}

\textbf{Comparison mode} enables cross-algorithm comparison at the objective-function level, since direct step-by-step alignment is impractical when each approach samples prompts in a different order. This makes the evolutionary relationships between algorithms visible without overwhelming non-expert users, remaining accessible to practitioners exploring algorithmic differences.
When users select a comparison algorithm, the two algorithms’ formulas are rendered vertically within each card, with color coding to mark removed, added, or modified components (\autoref{fig:algorithm_explainer}).
For example, users can compare the aggregation approaches of DAPO and Dr. GRPO to observe that Dr. GRPO removes DAPO's response length bias (\autoref{fig:algorithm_explainer}B).
The \textit{Algorithm Explainer} is laid out on a pan- and zoom-able canvas with automatic fitting, so longer formulas stay readable without crowding the screen, and cards other than the active objective remain collapsed until opened.
%
%
%
% The same JSON schema drives this comparison: components with matching user-defined IDs align their formula and text, while unmatched IDs are highlighted as added or removed.
The same JSON schema drives both comparison and extension: matching user-defined IDs align a component's formula and text while unmatched IDs are flagged as added or removed. Registering a new algorithm is likewise schema-only---its formula terms carry IDs that bind to logged fields, so real values populate the equation with no code changes.

\section{Usage Scenario}

We present two hypothetical \tool{} usage scenarios demonstrating its support for pedagogy and research: 
(1) an instructor and students explore a counterintuitive training step to build intuition for GRPO's group-relative reward dynamics (\autoref{sec:learning_grpo}); (2) an AI researcher compares DAPO and Dr. GRPO objectives to design a hybrid algorithm for her training data (\autoref{sec:algorithm_comparison}).

\subsection{Instructor Teaching GRPO via Training Behavior}
\label{sec:learning_grpo}

Professor Quinn is updating his graduate reinforcement learning course to showcase modern applications of RL to LLM fine-tuning. Knowing from prior semesters that students struggle to build intuition from mathematical notation alone, particularly for newer algorithms where the optimization behavior is not apparent from the objective function. Searching for an alternative to static lecture slides, he adopts \tool{} for his next class, focusing on GRPO to help students build intuition for how the algorithm determines which tokens are reinforced during training.

\setlength{\columnsep}{8pt}%
\setlength{\intextsep}{2pt}%
\begin{wrapfigure}{R}{0.15\textwidth}
    \centering
    \includegraphics[width=0.15\textwidth]{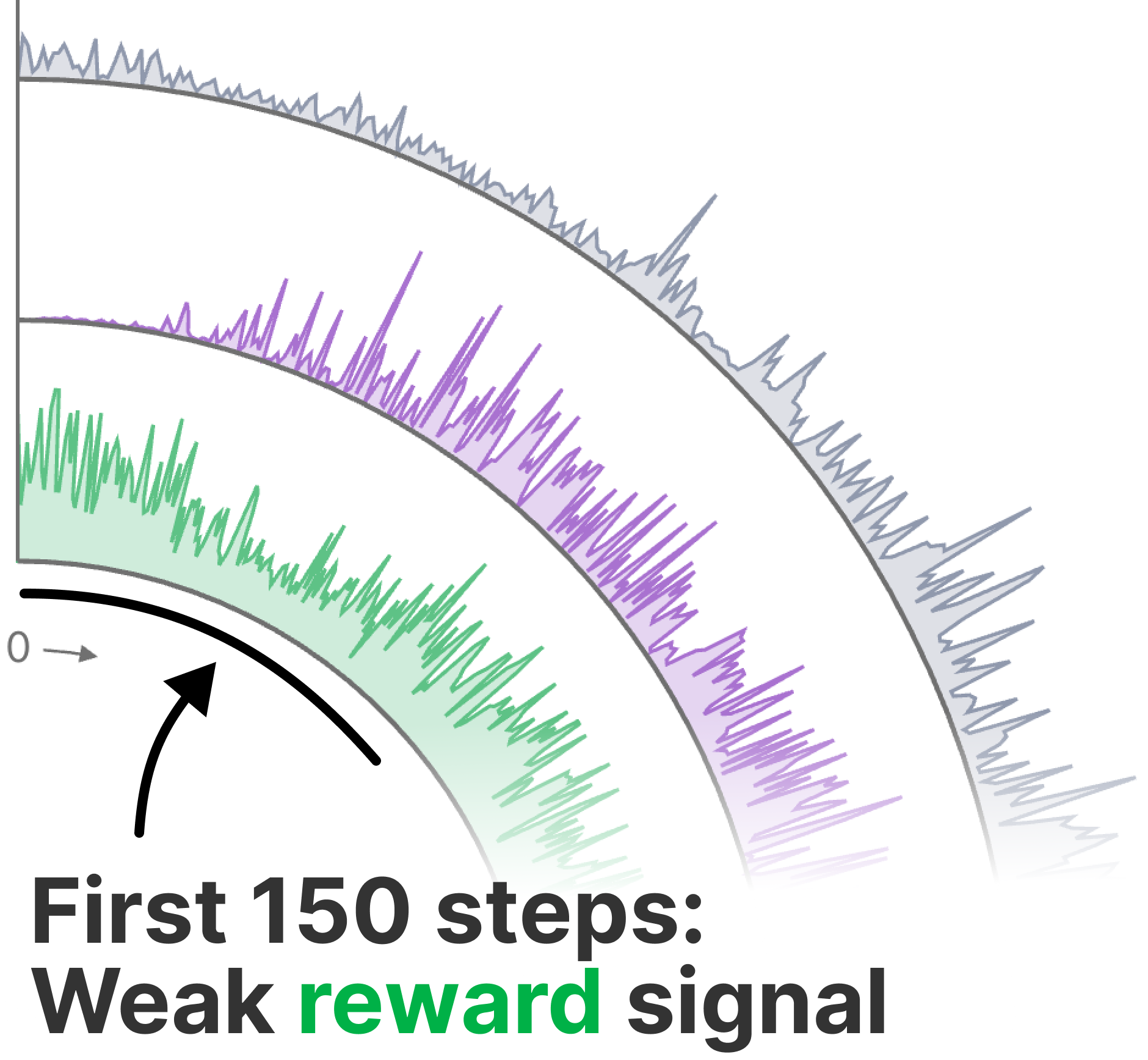}
\end{wrapfigure}

\paraheader{Surveying training dynamics with the Training Explorer.} 
During his next lecture, Quinn opens \tool{} and selects the GRPO algorithm from the navigation bar, loading the \textit{Training Explorer} view (\autoref{fig:teaser}A) for the MATH reasoning task.
From the metric selector, Quinn enables three rings on the radial plot across the 1,249 steps in the GRPO training logs.
\textcolor[HTML]{999999}{\textbf{Response length}}
(outermost ring) grows steadily as the model produces longer reasoning chains, 
\textcolor[HTML]{A377CB}{\textbf{policy clip ratio}} (middle ring) rises as the policy drifts, and \textcolor{green}{\textbf{reward}} (inner ring) shows a weak signal in the first 150 steps, with high step-to-step variance afterward. He uses this view to teach a key idea: meaningful training signal in GRPO accumulates through distributional shifts that the radial layout makes apparent, even if individual steps exhibit the step-to-step variance typical of policy gradient methods.

\paraheader{Investigating a counterintuitive training step with Step Inspector.} Quinn clicks step 242 in the radial plot, opening the \textit{Step Inspector} view (\autoref{fig:step_inspector}A), 
which displays the step's prompts and their response groups. He directs the class's attention to Prompt 2: 
\textit{``Two numbers have a difference of 3 and a sum of 31. What is the larger of the two numbers?''} 
Because of this problem's simplicity, all sampled responses arrive at the correct answer of 17. He asks the class to predict the token gradients of this response group; a student might expect a strong positive gradient from the intuition that high ``correctness'' should lead to a strong gradient signal. Quinn directs their attention to the response tokens themselves, which appear in white rather than the green reinforcement coloring students expected (\autoref{fig:step_inspector}A). The \textit{Step-level Objective} at the bottom of the panel reads {\setlength{\fboxsep}{1pt}\colorbox{pink!30}{$-0.323$}}, but none of this update comes from prompt 2's tokens. This is counterintuitive: every response is correct, reflecting a reward of \textcolor{green}{$1.00$}, but the model does not learn from this prompt at all.

\paraheader{Diagnosing the zero-gradient moment with the Algorithm Explainer.} To explain what the students just observed, Quinn opens the \textit{Algorithm Explainer} view by clicking one of the white tokens to reveal the per-token Advantage and Importance Ratio (\autoref{fig:step_inspector}B). 
For prompt 2, every response carries a non-zero Importance Ratio, but is scaled against an Advantage of \textcolor{softBlue}{$0.000$}. Returning to the \textit{Step Inspector} view, Quinn explains that the step-level objective of {\setlength{\fboxsep}{1pt}\colorbox{pink!30}{$-0.323$}} 
is driven by contributions from other prompts within the step, where the LLM produced a diverse group of responses. This reframes the students' intuition on the GRPO: correct responses are not automatically reinforced, and only responses correct \textit{relative to the group} provide useful updates to the policy.

\subsection{Researcher Comparing Algorithms for Training Run}
\label{sec:algorithm_comparison}

Riley is an AI researcher deciding between DAPO and Dr. GRPO for training a reasoning model on olympiad-level mathematics. She has implemented GRPO in OpenRLHF and monitored runs via Weights \& Biases~\cite{wandb}, but each technique frames its modifications relative to GRPO with its own notation, making direct DAPO vs. Dr. GRPO comparison difficult. She uses \tool{}'s \textit{Algorithm Explainer} to compare the two at the objective-function level before committing to a scaled training run.

\paraheader{Comparing objective functions side-by-side.} Riley opens \tool{}, selects DAPO from the algorithm navigation bar, and navigates to the \textit{Algorithm Explainer} view. She clicks Dr. GRPO into the comparison slot, rendering both objective functions side-by-side with color-coded modifications (\autoref{fig:algorithm_explainer}B). Curious how each algorithm handles aggregation, she expands the aggregation portion of the objective function and sees that Dr. GRPO replaces DAPO's implicit token weight with a constant max token length. 

\paraheader{Using \tool{} to determine task-specific suitability.} Riley hovers over DAPO's dynamic sampling constraint and reads that DAPO skips response groups where every response is correct or wrong. She connects this with her knowledge of the training data: as the model improves on easier olympiad problems, those prompts will contribute no gradient signal during training. Returning to the aggregation panel, Riley reads that DAPO's implicit token weight introduces a \textit{cross-group length bias} (\autoref{fig:algorithm_explainer}B), where a long response in a group of short responses dominates the group's gradient. Dr. GRPO's \textit{constant max token length} removes this bias, making it far better suited to the long chain-of-thought solutions in typical olympiad problems.

\paraheader{Integrating a new algorithm into \tool{}.} Riley concludes that neither algorithm will fully fits her needs: she wants DAPO's sampling efficiency \textit{and} Dr. GRPO's robust normalization scheme. Since dynamic sampling is a filtering step before the gradient update, she hypothesizes it could fold into Dr. GRPO's aggregation. Riley decides to prototype this hybrid approach, an insight enabled by \tool{}'s side-by-side comparison. Since \tool{} is open-source, she plans to contribute her hybrid algorithm and training logs back to the project.

\section{Conclusion}

We introduce \tool{}, to our knowledge the first interactive web-based tool that unifies the explanation of the RL fine-tuning algorithm family through three coordinated views. \tool{} connects training dynamics to the token-level objective computation,  making the algorithms' evolutionary relationships visible without requiring users to reconcile their differences. 
We open-source \tool{} as an extensible platform for the community to contribute new algorithms as the RL fine-tuning landscape rapidly evolves.

\acknowledgments{This work was supported in part by NSF \#2403297, 2502793, gifts from Google, Amazon, Meta, NVIDIA, Avast, Fiddler Labs, Bosch.}

\balance
\bibliographystyle{abbrv-doi-narrow}

\bibliography{template}
\end{document}